# Quasiparticle breakdown in a quantum spin liquid

Matthew B. Stone[1,3], Igor A. Zaliznyak[2], Tao Hong[3], Collin L. Broholm[3,4] and Daniel H. Reich[3]

[1]Condensed Matter Sciences Division, Oak Ridge National Laboratory, Oak Ridge, TN 37831, USA

[2]Physics Department, Brookhaven National Laboratory, Upton, NY 11973, USA

[3]Department of Physics and Astronomy, The Johns Hopkins University, Baltimore, Maryland 21218, USA

[4]National Institute of Standards and Technology, Gaithersburg, Maryland 20899, USA

**Much of modern condensed matter physics is understood in terms of elementary excitations, or quasiparticles – fundamental quanta of energy and momentum[1,2]. Various strongly-interacting atomic systems are successfully treated as a collection of quasiparticles with weak or no interactions. However, there are interesting limitations to this description: the very existence of quasiparticles cannot be taken for granted in some systems. Like unstable elementary particles, quasiparticles cannot survive beyond a threshold where certain decay channels become allowed by conservation laws – their spectrum terminates at this threshold. This regime of quasiparticle failure was first predicted for an exotic state of matter, super-fluid helium-4 at temperatures close to absolute zero – a quantum Bose-liquid where zero-point atomic motion precludes crystallization[1-4]. Using neutron scattering, here we show that it can also occur in a quantum magnet and, by implication, in other systems with Bose-quasiparticles. We have measured spin excitations in a two dimensional (2D) quantum-magnet, piperazinium hexachlorodicuprate (PHCC),[5] in which spin-1/2 copper ions form a non-magnetic quantum spin liquid**

**(QSL), and find remarkable similarities with excitations measured in superfluid $^4$He. We find a threshold momentum beyond which the quasiparticle peak merges with the two-quasiparticle continuum. It then acquires a finite energy width and becomes indistinguishable from a leading-edge singularity, so that excited states are not quasiparticles but occupy a wide band of energy. Our findings have important ramifications for understanding phenomena involving excitations with gapped spectra in many condensed matter systems, including high-transition-temperature superconductors[6].**

Although of all the elements only liquid helium fails to crystallize at $T = 0$, quantum liquids are quite common in condensed matter. Metals host electron Fermi liquids and superconductors contain Bose liquids of Cooper pairs. Trapped ultracold atoms can also form quantum liquids, and some remarkable new examples were recently identified among quantum spins in magnetic crystals[5,7-10]. The organo-metallic material PHCC is an excellent physical realization of a QSL in a 2D Heisenberg antiferromagnet (HAFM). Its $Cu^{2+}$ spins are coupled through a complex super-exchange network in the crystalline *a-c* plane forming an array of slightly skewed anisotropic *spin-½* ladders[10] coupled by frustrated interactions[5]. The spin excitations in PHCC have a spectral gap $\Delta_s \approx 1$ meV and nearly isotropic 2D dispersion in the (*h0l*) plane with a bandwidth slightly larger than $\Delta_s$. In the absence of a magnetic field, only short-range dynamic spin correlations typical of a liquid exist: the spin gap precludes long-range magnetic order down to $T = 0$. Here we explore magnetic excitations in PHCC via inelastic neutron scattering and compare the results with similar measurements in the quantum-fluid $^4$He, emphasizing the effects where the quasi-particle dispersion reaches the threshold for two-particle decay and interferes destructively with the continuum.

The properties of superfluid[4] $^4$He can be explained by considering Bose quasiparticles with a finite energy minimum (an energy gap) in their spectrum[1,2].

However, in a Bose quantum liquid, a spectral gap can produce an energy-momentum threshold where the quasiparticle description breaks down[1-3]. Beyond this threshold, single-particle states are no longer approximate eigenstates of the Hamiltonian and the quasiparticle spectrum terminates. Neutron scattering experiments in $^4$He indicate that the spectrum of quasiparticles (phonons) ends when the phonon is able to decay into two "rotons"[11-15]. These rotons are phonons with roughly quadratic dispersion that occur near the dispersion minimum, $\Delta \approx 0.74$ meV and wavevector $Q \approx 2$ Å$^{-1}$, *cf.* Fig. 1a. Spontaneous decays provide the only mechanism that destroys quasiparticles in $^4$He at $T = 0$. However, due to the high density of two-roton states, this decay path is so effective that instead of acquiring a finite lifetime, the quasiparticles simply cease to exist. Specifically, the single-particle pole is absent in the Green's function of $^4$He atoms for $Q > Q_c$, so that the quasiparticle spectrum does not continue beyond the threshold[1-3].

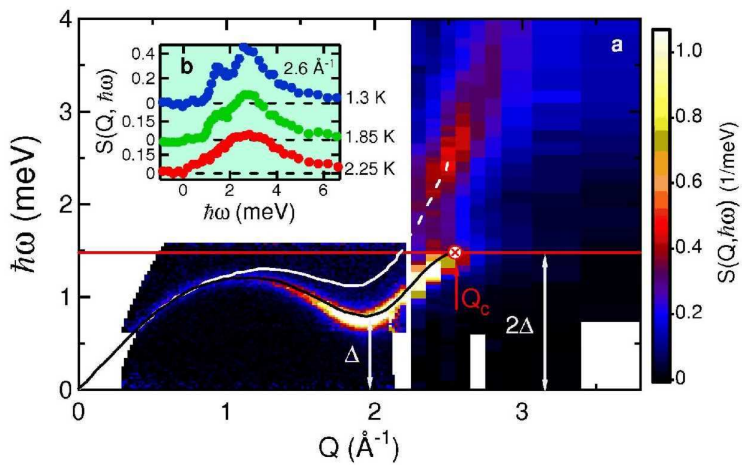

**Figure 1** Liquid helium excitation spectrum. **a** Excitation spectrum in $^4$He for 1.5 ≤ $T$ ≤ 1.8 K from inelastic neutron scattering measurements[13,16]. Solid black line is dispersion from Ref. 13; red circle with cross indicates spectrum termination point at $Q = Q_c$ and $\hbar\omega = 2\Delta$. White line is Feynman-Cohen bare dispersion in absence of decays[18], and horizontal red line at $\hbar\omega = 2\Delta$ shows onset of two-

roton states for $\hbar\omega \geq 2\Delta$. **b** Inset depicts excitations near termination point, at $Q$ = 2.6 Å$^{-1}$ ≈ $Q_c$, for several temperatures[13].

The excitation spectrum of superfluid $^4$He as probed by neutron scattering[13,16] is shown in Fig. 1a. One can see the roton minimum in the dispersion and the spectrum termination point at $Q_c \approx 2.6$ Å$^{-1}$. Near $Q_c$ the phonon hybridizes with two-roton excitations, its dispersion flattens, and spectral weight is transferred to the multiparticle continuum[13,15]. While a smeared maximum occurs at the leading edge of the continuum for $Q > Q_c$ and appears to continue the quasiparticle dispersion relation, it is instead ascribed to a two-roton bound state (resonance) resulting from roton-roton interactions[15,17]. Decays modify the "bare" Feynman-Cohen quasiparticle dispersion in $^4$He (white line in Fig 1a)[18]. Instead of terminating where it reaches the energy $2\Delta$, the quasiparticle spectrum is suppressed to lower energies at $Q \leq Q_c$, approaching the threshold $E = 2\Delta$ horizontally[3] (black line in Fig 1a).

The generality of the physics underlying quasiparticle breakdown in $^4$He suggests that similar effects may occur in other quantum liquids. The quasiparticle instability in $^4$He relies on the isotropic nature of the fluid: since the spectrum only depends on |**Q**|, the roton minimum produces a strong singularity in the density of states (DOS). For QSLs on a crystalline lattice, the DOS available for quasiparticle decays is enhanced by the absence of dispersion in certain directions that occurs in low dimensional systems (D<3) and in systems with competing interactions (frustration). Quasiparticle breakdown effects should thus be strongest in 1D QSLs, such as spin-1 chains with a spectral gap[19]. Though the term has not been used in this context, numerical work suggests that spectrum termination does occur in spin-1 HAFM spin chains[20,21]. Its observation through neutron scattering, however, is hindered by small scattering cross sections at the appropriate wavevectors. In the spin-1 chain system NENP scattering becomes undetectable when the single-particle excitation meets the non-interacting two-particle continuum[22], either due to decays or to a vanishing structure factor. While transformation of magnetic excitations from well-defined quasiparticles to a continuum was observed in the quasi-1D spin-1 HAFM $CsNiCl_3$, it is only seen as an onset of damping beyond a certain momentum threshold, well before the dispersion crosses the lower bound of the projected two particle continuum[23], which may be a result of inter-chain interactions.

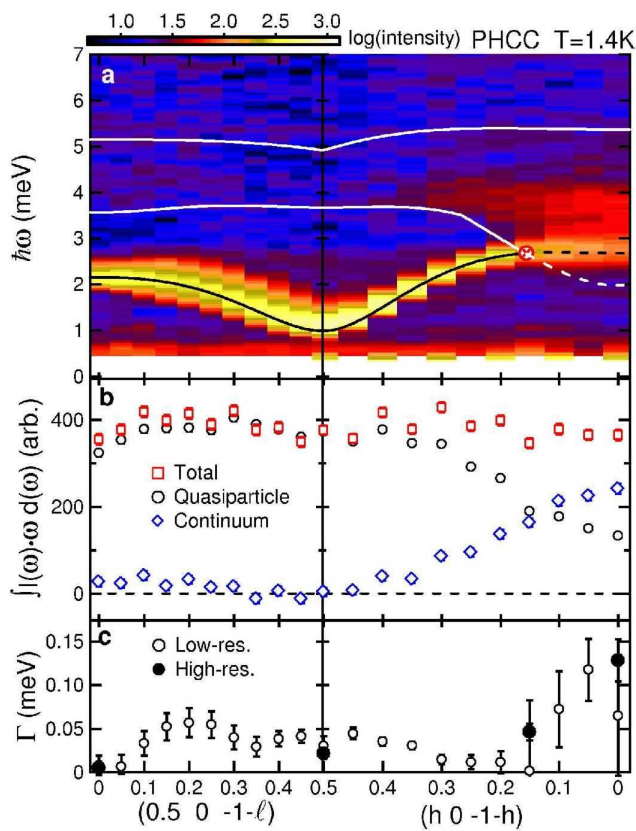

**Figure 2** Magnetic excitation spectrum at T=1.4 K in PHCC. **a** Background corrected intensity along the (½, 0, -1 - *l*) and (*h*, 0, *-1-h*) directions. A $\delta\hbar\omega$ = 0.25 meV running average was applied to each constant wave-vector scan, retaining the actual point density of the acquired data. Black line is previously determined single-magnon dispersion[5]. White lines are bounds of two-magnon continuum calculated from this dispersion. Red circle with cross indicates the point where the single particle dispersion relation intersects the lower bound of the two-particle continuum. **b** First frequency moment of measured scattering intensity integrated over different energy ranges. Red squares (total) correspond to 0.8 ≤ $\hbar\omega$ ≤ 5.5 meV, black circles (quasiparticle) to 0.8 ≤ $\hbar\omega$ ≤ 3 meV, and blue diamonds (continuum) to 3 ≤ $\hbar\omega$ ≤ 5.5 meV. **c** resolution corrected half width at half maximum of the lower energy peak throughout the range of wavevector transfer for high resolution (solid points) and low resolution (open points) data.

In contrast to the HAFM spin-1 chain, the structure factor of PHCC is favourable for probing the interaction of magnon quasiparticles with their two particle continuum. Its effects, however, could be less pronounced because the 2D DOS singularities are weaker. Prior measurements examined magnetic excitations in PHCC below ≈ 3 meV[5]. Here we present data for energies $\hbar\omega$ ≤ 7 meV and for wavevectors along the (½, 0, *l*) and (*h*, 0, -1-*h*) directions, elucidating both single- and multiparticle excitations in this 2D QSL. Data shown in Fig. 2a and selected scans shown in Fig. 3 demonstrate clear similarities to the spectrum of superfluid $^4$He. The one-magnon dispersion reaches the lower boundary of the two-magnon continuum, $\hbar\omega_{2m}(\mathbf{Q}) = \min_{\mathbf{q}}\{\hbar\omega(\mathbf{q}) + \hbar\omega(\mathbf{Q}-\mathbf{q})\}$, for $\mathbf{Q}_c = (h_c, 0, -1-h_c)$ with $h_c \approx 0.15$ near the magnetic Brillouin zone (BZ) boundary. The first frequency moment[24] integrated over different ranges of energy transfer shown in Fig. 2b reveals how oscillator strength is transferred from the quasi-particle excitation to the multiparticle continuum, in analogy to what is observed in $^4$He[13].

A change in the character of the excitation spectrum near $h_c$ is also apparent in Fig. 3, which shows the energy-dependent magnetic scattering for wavevectors along the (*h*, 0, -1-*h*) direction at $T \approx 1.4$ K << $\Delta_s$. For $h \geq 0.2$, Figs. 3a-c, there are two distinct contributions, a resolution-limited quasiparticle peak at lower energy and a broad feature with a sharp onset at higher energy, which we associate with the two-particle continuum. This continuum is well described by a square-root singularity above an energy threshold, typical for two-particle scattering governed by a diverging spectral density[20]. The threshold obtained from such data analysis is slightly higher than the calculated low-E boundary of the two-magnon continuum (white line in Fig. 2a), and is close to the lowest energy of two-particle states involving gap mode magnons with a diverging DOS. Alternatively, the shift could indicate magnon repulsion.

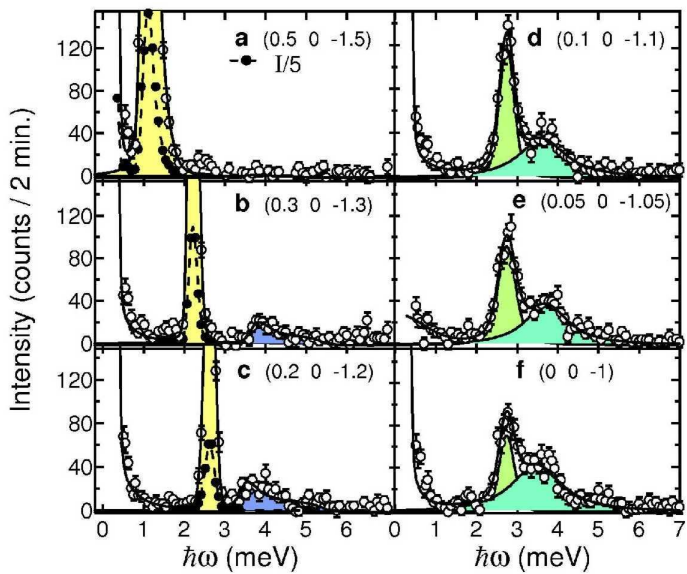

**Figure 3** Individual constant wavevector scans of PHCC along the ($h$, 0, -1-$h$) direction at $T$ = 1.4 K. Identical vertical scales emphasize variation in lineshape in vicinity of $h_c$. Solid lines are fits to single resonant mode (yellow shaded region) plus a higher energy continuum excitation (blue shaded region) convolved with the instrumental resolution function. For wavevectors $h \geq 0.2$, higher energy excitations are well represented by a two-particle continuum of the form $I = \dfrac{A}{\sqrt{(\hbar\omega)^2 - \varepsilon_1^2(Q)}} \Theta(\hbar\omega - \varepsilon_1(Q))\Theta(\varepsilon_2(Q) - \hbar\omega)$ with $\varepsilon_2(Q)$ defined by the calculated upper boundary of the two-particle continuum (white line in Fig. 2a); $\varepsilon_1(Q)$ and $A$ were refined by the least square fitting. For $h \leq 0.15$ this description fails and the spectrum is fitted by two superimposed DHO spectra,

$$I = \frac{\Gamma}{\pi}\left(\frac{1}{\Gamma^2 + (\hbar\omega - \hbar\omega_0)^2} - \frac{1}{\Gamma^2 + (\hbar\omega + \hbar\omega_0)^2}\right)$$ (green shaded regions). The Gaussian representing elastic incoherent nuclear scattering is also included at all wavevectors. Dashed lines and solid symbols in panels a-c show data on a one-fifth intensity scale.

For $h \leq 0.15$ the quasiparticle peak joins the continuum to form a complex spectral feature that extends from 2.5 to 4.5 meV (Figs. 3d-f). We parameterize this spectrum by the overlapping response of two damped harmonic oscillators (DHO). The onset of scattering occurs well above the lowest energy for two non-interacting magnons (dashed white line in Fig. 2a), which indicates significant interactions. While the lower energy peak that appears to continue the quasiparticle dispersion in PHCC carries more spectral weight than the corresponding resonance at $Q > Q_c$ in superfluid $^4$He, it also has a measurable energy width as quantified in Fig. 2c. This demonstrates that a decay mechanism abruptly becomes accessible to the low energy excitation for $h \leq 0.15$. The width increases towards the BZ boundary, $h = 0$, where the peak at the leading edge can be described by a non-quasiparticle square root singularity as used for the continuum at $h \geq 0.2$, or as an unstable non-dispersive resonance below the continuum.

The temperature dependence of scattering in $^4$He for $Q$ between 2.4 and 2.6 Å$^{-1}$, Fig. 1b,[13,14] provides additional evidence of quasiparticle spectrum breakdown. Data in PHCC for $\mathbf{Q} = (0.15, 0, -1.15)$ where the one and two magnon states converge shown in Figs. 4 e-h, similarly indicate that proximity to the two-particle continuum enhances thermal damping: the peak whose energy is approximately 20 K is severely broadened already at T=10 K (Fig. 4 f). Its thermal broadening resembles that of the $\mathbf{Q} = (0.5, 0, -1.5)$ gap mode, which is shown in Fig. 4 a-d. This differs from observations in copper nitrate, a 1D QSL with weak dispersion where the one-magnon band lies well below the two-magnon continuum and spectrum termination cannot occur[25]. Temperature-induced damping in that case is stronger for the lower-energy gap mode than for quasiparticles at the top of the dispersion curve, i.e. heating mainly affects energy levels that become thermally populated. For PHCC, damping near the top and bottom of the band is governed by the same thermal population (inset Fig. 4a), consistent with the idea that high-energy excitations decay into gap-mode quasiparticles. As their thermal population increases, the probability of stimulated emission by the high-energy excitations also grows.

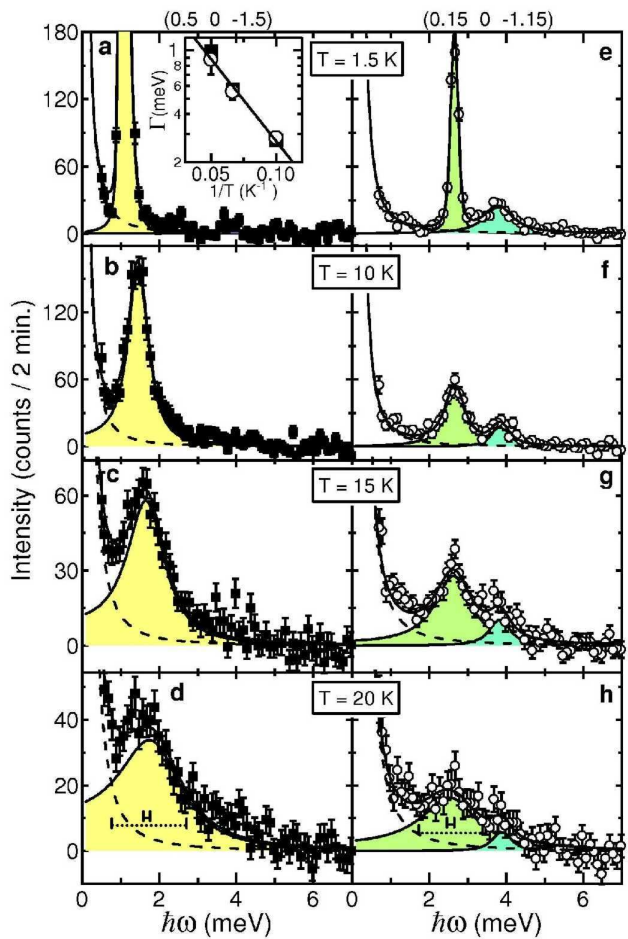

**Figure 4** Temperature dependent energy spectra for PHCC at $Q$ = (0.5, 0, -1.5) (**a-d**) and $Q$ = (0.15, 0, -1.15) (**e-h**). Solid lines for $T$ = 1.5 K in **a** and **b** are fits as described in Fig. 3. Solid lines for $T \geq 10$ K are fits to the following response function satisfying detailed balance constraint

$$S(Q,\omega) = \frac{\Gamma}{\pi}\left(\frac{1}{\Gamma^2 + (\hbar\omega - \hbar\omega_0)^2} - \frac{1}{\Gamma^2 + (\hbar\omega + \hbar\omega_0)^2}\right)\frac{1}{1-\exp(-\beta\hbar\omega)}.$$

Temperature dependence of the relaxation rate, $\Gamma$, for the lower energy peak at both wavevectors is shown in inset to **a**. Line corresponds to exponentially activated behaviour with $\Delta=2.0$ meV. Coloured areas below peaks indicate the assignment of different contributions to the spectra. Dashed lines indicate incoherent elastic nuclear scattering. Solid (dashed) horizontal bars in frames **d** and **h** indicate resolution (width of the low energy peak).

In summary, quasiparticle spectrum termination as seen in superfluid $^4$He can also occur in other condensed matter systems, quantum magnets in particular. Dramatic changes that we observe in the spectrum of magnetic excitations in PHCC provide compelling evidence for its existence in the 2D QSL. The termination point is marked by rapid transfer of intensity from the magnon peak to the continuum at higher energies and by an abrupt appearance of damping. Although in PHCC the damped peak at the leading edge of magnetic scattering carries more intensity than the analogous peak in superfluid $^4$He, the line-shape and temperature dependence of post threshold excitations in these two very different quantum liquids are remarkably similar.

Quasiparticles are ubiquitous in nature ranging from phonons, magnons, rotons[1-3], magnetorotons[26] and heavy electrons and holes in condensed matter physics to the quasiparticles of the quark gluon plasma and the various unstable particles and resonances in the standard model of particle physics[27]. Rarely, however, do experiments offer as detailed a view of quasiparticle decay as the present results in a 2D organo-metallic spin liquid. Our findings show that an analysis of excitations in terms of quasiparticles with a well-defined dispersion relation can be at fault beyond a certain energy-momentum threshold where the quasiparticles break down. This has important implications for a variety of condensed matter systems, in particular for other QSLs such as lamellar copper oxide superconductors, where spin excitations above a gap are considered as possible mediators of electron pairing and high-temperature superconductivity[28].

**Methods**

Neutron scattering measurements of PHCC were performed using the SPINS cold neutron triple axis spectrometer at the NIST Center for Neutron Research. Four deuterated PHCC crystals[5] with a total mass of 7.5 g were co-aligned to within 1º. Energy scans were acquired by varying the incident beam energy for fixed monitor counts in a low-efficiency detector between the pyrolytic graphite (PG (002)) monochrometer and the sample. A 138' radial collimator was used between the sample and a horizontally focusing PG (002) analyzer with an angular acceptance of 5º horizontally and 6º vertically. A cooled Be filter was in place after the sample. Measurements in Fig. 4 employed an additional PG filter before the sample. Data in Figs. 2 and 3 [Fig. 4] were acquired with 5 [3.7] meV fixed final energy. Projected full width at half maximum energy resolution of these configurations at $\hbar\omega = 0$ is 0.18 meV and 0.11 meV respectively. A wavevector independent fast-neutron background was measured by shielding the analyzer entrance with cadmium. A wavevector-dependent thermal neutron background arising predominantly from incoherent phonon scattering was measured at T=100 K and scaled using the thermal detailed balance factor for use as a low-temperature non-magnetic background. These backgrounds were subtracted from all data presented.

**Acknowledgements** We acknowledge discussions with L. Passell, J. Tranquada, A. Abanov, M. Zhitomirsky, A. Tsvelik, A. Chitov, and M. Swartz. Work at BNL and ORNL was supported by the Office of Science, U.S. Department of Energy, under contracts DE-AC02-98CH10886 and DE-AC05-00OR2272. Work on SPINS and at JHU was supported by the U.S. National Science Foundation. We are grateful to B. Fåk for permission to reproduce the results of his measurements on $^4$He in Fig. 1.

**Correspondence** and requests for materials should be addressed to I.Z. (zaliznyak@bnl.gov).